\documentclass[amsmath,amssymb,aps,preprintnumbers,twocolumn]{revtex4-1}
\pdfoutput=1
\usepackage{epsf}

\usepackage{epsfig,graphics}
\usepackage {graphicx}
\usepackage {epsfig}
\usepackage {subfigure}
\usepackage {tabularx} 
\usepackage{rotate}	
\usepackage{slashed}

\usepackage{amsmath}
\usepackage{amsfonts}
\usepackage{amssymb}
\usepackage{graphicx}

\newcommand{\bmat}{\left(\begin{array}}
\newcommand{\emat}{\end{array}\right)}

\def\yzero{\smash{\hbox{$y\kern-4pt\raise1pt\hbox{${}^\circ$}$}}}

\def\beq{\begin{equation}}
\def\eeq{\end{equation}}
\def\beqa{\begin{eqnarray}}
\def\eeqa{\end{eqnarray}}

\def\-{\hphantom{-}}

\def\s2{\frac{1}{\sqrt2}}

\def\beq{\begin{equation}}
\def\eeq{\end{equation}}
\def\beqa{\begin{eqnarray}}
\def\eeqa{\end{eqnarray}}

\def\IF{\relax{\rm I\kern-.18em F}}
\def\II{\relax{\rm I\kern-.18em I}}
\def\IP{\relax{\rm I\kern-.18em P}}
\def\IC{\relax\hbox{\kern.25em$\inbar\kern-.3em{\rm C}$}}
\def\IR{\relax{\rm I\kern-.18em R}}

\def\Dsl{\,\raise.15ex\hbox{/}\mkern-13.5mu D} 
\def\IZ{Z\kern-.4em  Z}

\catcode`\@=11   
\newdimen\@rotdimen
\newbox\@rotbox  

\def\@vspec#1{\special{ps:#1}}
\def\@rotstart#1{\@vspec{gsave currentpoint currentpoint translate
   #1 neg exch neg exch translate}}
\def\@rotfinish{\@vspec{currentpoint grestore moveto}}
\def\@rotr#1{\@rotdimen=\ht#1\advance\@rotdimen by\dp#1%
   \hbox to\@rotdimen{\hskip\ht#1\vbox to\wd#1{\@rotstart{90 rotate}%
   \box#1\vss}\hss}\@rotfinish}
\def\@rotl#1{\@rotdimen=\ht#1\advance\@rotdimen by\dp#1%
   \hbox to\@rotdimen{\vbox to\wd#1{\vskip\wd#1\@rotstart{270 rotate}%
   \box#1\vss}\hss}\@rotfinish}%
\def\@rotu#1{\@rotdimen=\ht#1\advance\@rotdimen by\dp#1%
   \hbox to\wd#1{\hskip\wd#1\vbox to\@rotdimen{\vskip\@rotdimen
   \@rotstart{-1 dup scale}\box#1\vss}\hss}\@rotfinish}%
\def\@rotf#1{\hbox to\wd#1{\hskip\wd#1\@rotstart{-1 1 scale}%
   \box#1\hss}\@rotfinish}%
\def\rotate{\@ifnextchar[{\@rotate}{\@rotate[l]}}
\def\@rotate[#1]#2{\setbox\@rotbox=\hbox{#2}\@nameuse{@rot#1}\@rotbox}

\catcode`\@=12

\begin{document}

\preprint{IFT-UAM/CSIC-17-066}
\preprint{MPP-2017-158}

\title{Constraining the EW Hierarchy \\
from the Weak Gravity Conjecture}
\author{Luis E. Ib\'a\~nez$^1$}
\author{ V\'ictor Mart\'in-Lozano$^2$}
\author{Irene Valenzuela$^{3,4}$}
\address{$^1$Departamento de F\'{\i}sica Te\'orica and Instituto de F\'{\i}sica Te\'orica  UAM-CSIC, Universidad Aut\'onoma de Madrid, Cantoblanco, 28049 Madrid, Spain\\
$^2$Bethe Center for Theoretical Physics \& Physikalisches Institut der Universit\"{a}t Bonn, Nu{\ss}allee 12, 53115, Bonn, Germany\\
$^3$Max Planck Institute for Physics, F\"ohringer Ring 6, 80805 Munich, Germany \\
$^4$Institute for Theoretical Physics and Center for Extreme Matter and Emergent Phenomena, 
      Utrecht University, Leuvenland 4, 3584 CE Utrecht, The Netherlands}
\begin{abstract}
 In this addendum we complement the remarks made in ref[1] constraining the value of the cosmological constant  $\Lambda_4$
in terms of neutrino masses. Those were based on a sharpened version of the Weak Gravity Conjecture as applied to 
compactifications of the SM to lower dimensions. We argue that the same line of reasoning implies that for fixed values 
of  $\Lambda_4$ and the Yukawa coupling of the lightest neutrino $Y_{\nu_1}$, the EW scale is bounded above. This is a trivial
consequence of neutrino masses depending on the Higgs vev.  In the case of massive 
Majorana neutrinos with a see-saw mechanism associated to a large scale  $M\simeq 10^{10-14}$ GeV and $Y_{\nu_1}\simeq 10^{-3}$, one obtains 
that the EW scale cannot exceed $M_{EW}\lesssim 10^2-10^4$ GeV. From this point of view, the delicate fine-tuning required to get a small EW scale would be a mirage, since parameters yielding higher EW scales would be in the {\it swampland} and would not count as possible consistent theories. This would bring a new perspective into the issue of the EW hierarchy.

\end{abstract}
\maketitle

{\it Introduction}.
In ref.\cite{Ibanez:2017kvh} we discussed in detail  how a  {\it sharpened} version \cite{Ooguri:2016pdq} of the Weak Gravity Conjecture (WGC) \cite{WGC} leads
to constraints both on the cosmological constant $\Lambda_4$  and neutrino masses. This sharpened WGC implies that no theory with an AdS 
non-SUSY stable vacuum can be embedded into a consistent theory of quantum gravity.  If one considers the compactification of the Standard Model (SM) plus 
Einstein's gravity down to 3D or 2D dimensions it turns out that, depending on the value of neutrino masses, AdS vacua may appear
\cite{ArkaniHamed:2007gg,Fornal:2011tw}.  The dependence on neutrino
masses in 3D or 2D  arises because in the infrared, below the scale of the electron mass, they give the leading (positive) contribution to the Casimir potential, 
to be added to the (negative) contribution of the only massless fields, the photon and the graviton. 
The absence of an AdS vacuum imposes an upper bound on the neutrino masses in terms of the 4D cosmological constant. Via the dependence of neutrino masses on the Higgs vev, this bound is translated into un upper bound for the EW scale, as we proceed to explain in the following.

{\it Upper bound on the EW scale}. 
The essential ingredient  to minimally avoiding  3D,2D  AdS vacua  is having 4 fermionic degrees of freedom sufficiently light (lighter than $\simeq \Lambda_4^{1/4}$) so
as to cancel the negative contribution coming from the photon and graviton,  before the radion potential becomes negative,  as the compact radii decrease.   It is then clear that,
{\it for a fixed value of $\Lambda_4$},  the mass of these lightest fermionic degrees of freedom is bounded from above. This may be seen e.g. 
in figs. $5$ and $6$ in  \cite{Ibanez:2017kvh}. 
In the case of Majorana neutrinos in addition to the ligthest neutrino an additional Weyl fermion state lighter than $10^{-3}$ eV must also be added if we want
to avoid AdS vacua. But again
one observes in fig.$12$ in  \cite{Ibanez:2017kvh} that there is an upper bound on the mass of the lightest neutrino (both in normal neutrino  hierachy (NI) and inverted hierarchy (IH)).
Similar results are obtained in compactifications to 2D.
\begin{figure}[t]
	\begin{center}
        \includegraphics[scale=0.33]{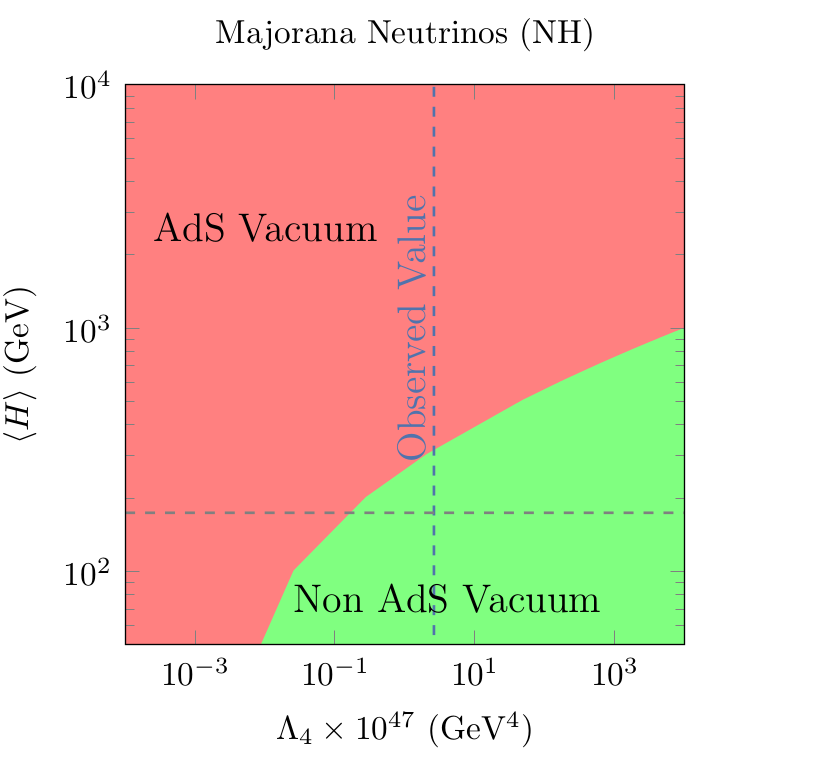}
		\caption{\footnotesize  Constraints on the EW scale and the cosmological constant for the case of Majorana neutrinos and normal hierarchy, in the presence of an additional Weyl fermion of mass $m_\chi=10^{-3}$ eV.  We have assumed $Y=10^{-3}$ and $M=10^{10}$ GeV.}
		\label{majorana}
	\end{center}
\end{figure}

If neutrinos are Majorana one sees from table 4 in  \cite{Ibanez:2017kvh} that $m_{\nu_1}\lesssim  5(1)\times 10^{-3}$ eV$\sim  2(0.4)\times \Lambda_4^{1/4}$
for NI (IH) respectively. If the lightest neutrino Majorana mass is induced from a standard see-saw mechanism one obtains (e.g for NI)
\footnote{Of course, one only obtains a useful bound if the lightest neutrino has non-zero mass.}
\beq
\frac {(Y_{\nu_1}<H>)^2}{M} \ \lesssim\ 2\times \Lambda_4^{1/4} \ \longrightarrow \  \\
<H> \ \lesssim  \ \frac {\sqrt{2}}{Y_{\nu_1}} \sqrt{M \Lambda_4^{1/4}}  \ .
\label{hierarchy}
\eeq 
where $M$ is the scale of lepton number violation in the see-saw.
Thus one gets the interesting conclusion that, for a  given fixed  c.c. $\Lambda_4$ and fixed Yukawa coupling, the EW scale  is bounded above by
the geometric mean of the cosmological constant scale and the lepton number violation scale $M$. Thus, e.g. for $Y_{\nu_1}\simeq 10^{-3}$ and
$M\simeq 10^{10}-10^{14}$ GeV, one gets $<H> \lesssim  10^2-10^4$ GeV.  Larger EW scales would yield (for fixed Yukawa) too large lightest neutrino mass and
AdS vacua would be generated. In other words, consistency with quantum gravity requires that a very small 4D cosmological constant should come accompanied by a big hierarchy between the EW scale and $M$. 
In figure \ref{majorana} we depicted the constraints on the EW scale (parametrised by the Higgs vev) and the 4D cosmological constant for fixed $Y=10^{-3}$ and $M=10^{10}$ GeV, leading to the aforementioned upper bound on the EW scale. Similar results apply for the case of inverted neutrino mass hierarchy.

\begin{figure}[t]
	\begin{center}
        \includegraphics[scale=0.33]{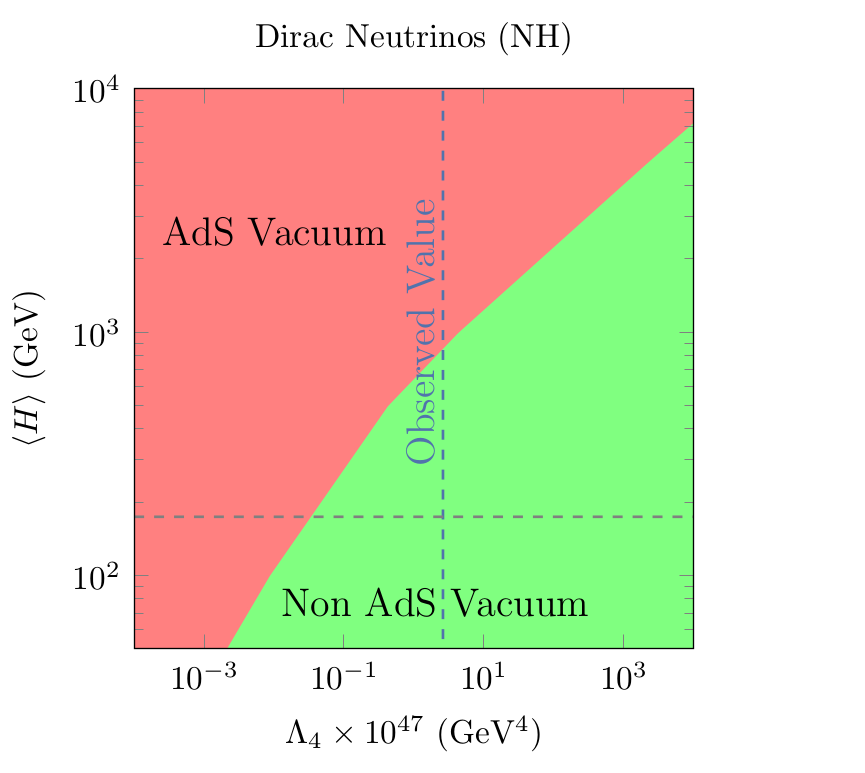}
		\caption{\footnotesize Constraints on the EW scale and the cosmological constant for the case of Dirac neutrinos and normal hierarchy. We have assumed a Yukawa coupling $Y=10^{-14}$.}
		\label{dirac}
	\end{center}
\end{figure}

In the case of Dirac neutrinos one rather gets 
$<H> \lesssim   1.6(0.4) \Lambda_4^{1/4} Y_{\nu_1}^{-1}$ for NI(IH). Now,  for fixed Yukawa coupling the EW scale is again bounded above by the 4D cosmological constant. In the Dirac case, 
though, the Yukawa
coupling needs to be extremely small to match the scale of observed neutrino masses
\footnote{In the case of Dirac neutrinos, one can also apply the argument in the opposite direction, 
 to explain why at least one of the neutrinos has a Yukawa $\lesssim 10^{-14}$.
Indeed {\it for fixed $\Lambda_4$ and EW scale}, one lightest neutrino  with a Yukawa coupling $\lesssim 10^{-14}$ would be enough  to
avoid the existence of $3D,2D$ AdS vacua. However the other two neutrino generations would not be constrained from such arguments.}. But again, the smallness of the cosmological constant implies in turn a small EW scale in order to be consistent with quantum gravity. This relation is shown in figure \ref{dirac} for fixed Yukawa coupling $Y=10^{-14}$.

{\it Discussion}. From the point of view a low energy field theorist the smallness of the EW scale is surprising because there is apparently nothing preventing 
the Higgs mass to grow up to the UV cut-off scale. That is the hierarchy problem. If that  huge UV mass squared is negative,  that would give rise to EW breaking close to the UV scale. 
We now see that,  from the WGC point of view  here considered,  that situation would not be possible (for fixed $\Lambda_4$) because AdS vacua would then be generated
at 3D and 2D compactifications.  The other option is having a positive UV scale mass for the Higgs, i.e., no Higgs at low energies at all.  That situation 
turns out to be also inconsistent with the WGC.  Indeed, starting with the SM with just fermions, gauge bosons and no Higgs, the theory has a global
accidental $U(6)_R\times U(6)_L$ symmetry in the quark sector. Once QCD condensation takes place, the symmetry is broken to the diagonal $U(6)$ and
36 Goldstone bosons appear. Out of those 3 are swallowed by the $W^\pm$ and $Z$ bosons. These large number of bosons outnumbers the 
massless leptonic  degrees of freedom  which are 18 or 24 if neutrinos are Dirac. This makes that again an AdS vacuum develops. 

From the present perspective the Higgs scale is small compared to the UV scale because of the smallness of the c.c. With values of $\Lambda_4$ 
as observed in cosmology, and reasonable non-vanishing lightest neutrino Yukawa, scales larger than the measured EW scale would yield theories with 3D,2D AdS vacua.  From the  Wilsonian effective field
theory point of view the smallness of the Higgs scale looks like a tremendous fine-tuning. However such a fine-tuning would be a mirage since 
parameters yielding higher Higgs mass scales or vevs cannot be embedded into a consistent theory of quantum gravity and hence do not count 
as possible consistent theories.

As discussed in \cite{Ibanez:2017kvh}, our results rely on the assumption that the lower dimensional vacua obtained from compactifying the SM to 3D or 2D are stable. Unfortunately, this issue cannot be addressed without a better understanding of the embedding of the SM into a possible landscape of vacua and the tunneling transitions between them. However, they exemplify how consistency with quantum gravity can have important implications on IR physics and change our conception about presumed \emph{fine-tuning} problems. This can bring a new perspective into the issue of the EW hierarchy and other issues of \emph{naturalness} in nature.

\begin{acknowledgments}

The work of L.E.I   has been supported by the ERC Advanced Grant SPLE under contract ERC-2012-ADG-20120216-320421, 
and the  spanish AEI and  the EU FEDER through the projects FPA2015-65480-P and FPA2016-78645-P, 
  as well as  the grant SEV-2012-0249 of the ``Centro de Excelencia Severo Ochoa" Programme.  V.M.L 
  acknowledges support of the Consolider MULTIDARK project CSD2009-00064, the SPLE ERC project  and the
  BMBF under project 05H15PDCAA. I.V. is suported by a grant of the Max Planck society.

\end{acknowledgments}



\begin{thebibliography}{2}%
\makeatletter
\providecommand \@ifxundefined [1]{%
 \@ifx{#1\undefined}
}%
\providecommand \@ifnum [1]{%
 \ifnum #1\expandafter \@firstoftwo
 \else \expandafter \@secondoftwo
 \fi
}%
\providecommand \@ifx [1]{%
 \ifx #1\expandafter \@firstoftwo
 \else \expandafter \@secondoftwo
 \fi
}%
\providecommand \natexlab [1]{#1}%
\providecommand \enquote  [1]{``#1''}%
\providecommand \bibnamefont  [1]{#1}%
\providecommand \bibfnamefont [1]{#1}%
\providecommand \citenamefont [1]{#1}%
\providecommand \href@noop [0]{\@secondoftwo}%
\providecommand \href [0]{\begingroup \@sanitize@url \@href}%
\providecommand \@href[1]{\@@startlink{#1}\@@href}%
\providecommand \@@href[1]{\endgroup#1\@@endlink}%
\providecommand \@sanitize@url [0]{\catcode `\\12\catcode `\$12\catcode
  `\&12\catcode `\#12\catcode `\^12\catcode `\_12\catcode `\%12\relax}%
\providecommand \@@startlink[1]{}%
\providecommand \@@endlink[0]{}%
\providecommand \url  [0]{\begingroup\@sanitize@url \@url }%
\providecommand \@url [1]{\endgroup\@href {#1}{\urlprefix }}%
\providecommand \urlprefix  [0]{URL }%
\providecommand \Eprint [0]{\href }%
\providecommand \doibase [0]{http://dx.doi.org/}%
\providecommand \selectlanguage [0]{\@gobble}%
\providecommand \bibinfo  [0]{\@secondoftwo}%
\providecommand \bibfield  [0]{\@secondoftwo}%
\providecommand \translation [1]{[#1]}%
\providecommand \BibitemOpen [0]{}%
\providecommand \bibitemStop [0]{}%
\providecommand \bibitemNoStop [0]{.\EOS\space}%
\providecommand \EOS [0]{\spacefactor3000\relax}%
\providecommand \BibitemShut  [1]{\csname bibitem#1\endcsname}%
\let\auto@bib@innerbib\@empty
\bibitem [{Note1()}]{Note1}%
  \BibitemOpen
  \bibinfo {note} {Of course, one only obtains a useful bound if the lightest
  neutrino has non-zero mass.}\BibitemShut {Stop}%
\bibitem [{Note2()}]{Note2}%
  \BibitemOpen
  \bibinfo {note} {In the case of Dirac neutrinos, one can also apply the
  argument in the opposite direction, to explain why at least one of the
  neutrinos has a Yukawa $\lesssim 10^{-14}$. Indeed {\protect \it for fixed
  $\Lambda _4$ and EW scale}, one lightest neutrino with a Yukawa coupling
  $\lesssim 10^{-14}$ would be enough to avoid the existence of $3D,2D$ AdS
  vacua. However the other two neutrino generations would not be constrained
  from such arguments.}\BibitemShut {Stop}%
\end{thebibliography}%


\begin{thebibliography}{99}


\bibitem{Ibanez:2017kvh}
  L.~E.~Ib\'a\~nez, V.~Mart\'{\i}n-Lozano and I.~Valenzuela,
  ``Constraining Neutrino Masses, the Cosmological Constant and BSM Physics from the Weak Gravity Conjecture,''
  arXiv:1706.05392 [hep-th].
  
  
  \bibitem{Ooguri:2016pdq}
H.~Ooguri and C.~Vafa,
``Non-supersymmetric AdS and the Swampland,''
arXiv:1610.01533 [hep-th].



\bibitem{WGC}
  C.~Vafa, ``The String landscape and the swampland,''
  hep-th/0509212\\
  N.~Arkani-Hamed, L.~Motl, A.~Nicolis and C.~Vafa,
  ``The String landscape, black holes and gravity as the weakest force,''
  JHEP {\bf 0706} (2007) 060
  [hep-th/0601001]\\
  H.~Ooguri and C.~Vafa,
  ``On the Geometry of the String Landscape and the Swampland,''
  Nucl.\ Phys.\ B {\bf 766}, 21 (2007)
  [hep-th/0605264].



\bibitem{ArkaniHamed:2007gg}
  N.~Arkani-Hamed, S.~Dubovsky, A.~Nicolis and G.~Villadoro,
  ``Quantum Horizons of the Standard Model Landscape,''
  JHEP {\bf 0706} (2007) 078
  [hep-th/0703067 [HEP-TH]].

\bibitem{Fornal:2011tw}
  B.~Fornal and M.~B.~Wise,
  ``Standard model with compactified spatial dimensions,''
  JHEP {\bf 1107} (2011) 086
  [arXiv:1106.0890 [hep-th]]\\
   J.~M.~Arnold, B.~Fornal and M.~B.~Wise,
  ``Standard Model Vacua for Two-dimensional Compactifications,''
  JHEP {\bf 1012} (2010) 083
  [arXiv:1010.4302 [hep-th]].


\end{thebibliography}
\end{document}